# Observations of Time Delays in Gravitational Lenses from Intensity Fluctuations: The Coherence Function

Ermanno F. Borra,
Centre d'Optique, Photonique et Laser,
Département de Physique, Université Laval, Québec, Qc, Canada G1K 7P4
(email: borra@phy.ulaval.ca)

**ABSTRACT**

      Measurements of the spectrum of the fluctuations of the output current of the quadratic detector of a telescope can be used to find unresolved astronomical gravitational lenses and determine time delays between their image components. These time delays can be used for astronomical studies. The spatial correlation coefficient of a source is an important parameter that quantifies the loss of contrast, caused by the extendedness of the source, in the spectral modulation of the intensity fluctuations. This work shows that the correlation coefficient must not be evaluated at the frequency of observation, but must instead be evaluated at the much lower beat frequencies of the spectrum of the fluctuations. This opens up a powerful novel technique to find unresolved gravitational lenses and to study the lensing event and the source.
.

## 1. INTRODUCTION

In general, when a distant source is gravitationally lensed by an intervening massive object, two images of the source are generated. The two beams from the source to the observer have different optical paths and therefore, if recombined in the focal plane of a detector, can interfere. Peterson & Falk (1991) have proposed to use interference effects in gravitational lenses to detect weak gravitational lenses and measure the time delays between the components. Classical interferometric techniques are however limited to the very small time delays generated by lensing objects having masses substantially lower than a solar mass. Motivated by the fact that the time delays among the images of a gravitational lens could be exploited for a variety of studies (Blandford & Naryan 1992), Borra (1997) pointed out that, due to interference effects, the spectral distributions of gravitational lenses carry signatures that could be used to identify unresolved lenses. Time delays of essentially arbitrarily length could be measured by analyzing the intensity signal, adapting a technique first used by Alford & Gold (1958), hereafter referred to as the AG effect, to measure the speed of light. However, Borra (1997) ended in a pessimistic note by concluding that spatial coherence effects would, in practice, make it difficult to use the technique in most astronomical lenses. This is because Borra (1997) made the usual assumption, valid for classical interferometry, that the spatial correlation coefficient, that describes the loss of contrast in the spectral modulation due to the AG effect, must be evaluated at the frequency of observation. On the basis of this frequency dependence, Borra (1997) concluded that, in practice, the technique had limited applications. In this article we do not make the same assumption but, instead, reconsider the physics of the problem from first principles and make a new estimate of the spatial coherence function. We then reconsider the application of the technique to gravitational lenses.

## 2. PHYSICAL PRINCIPLES

Figure 1 shows a layout of the gravitational lens interferometer. The gravitational lens is simply modeled by a Young interferometer having equal strength beams. If we use the classical simplified geometrical model for the two components of a gravitational lens (Refsdal 1964, Press & Gunn 1973), examination of the figure that describes the model shows that it can be approximately modeled by a Young interferometer that introduces time delays between two interfering beams. The gravitational lens can thus be modeled by a Young interferometer having a separation *a* between the slits which determines the time delay. The impact parameter *a* depends therefore on the mass of the lensing object. The unresolved image of the lens is observed with a conventional telescope that uses a conventional quadratic detector. The current from the detector is then sent to a wave analyzer that obtains the spectrum of the current fluctuations.

The technique is based on the experiment by Alford & Gold (1958), hereafter referred to as the AG effect, to measure the speed of light. The AG experiment was carried out with a set up that is essentially a Young interferometer. The AG effect may look strange but is fully established experimentally and has firm theoretical foundations (Givens 1961, Mandel 1962). To understand the AG effect one must, firstly, understand that interfering beams that have an optical path difference that considerably exceeds the temporal coherence length of the beams can generate a spectrally modulated beam; secondly that the electromagnetic waves in a primary spectrum beat among themselves to generate a lower frequency beat spectrum; thirdly that the spectral distribution of beats, hereafter referred to as beat spectrum, can be obtained from intensity measurements with a square-law detector.

While classical interferometry is described in textbooks (e.g. Klein & Furtak, 1986), it is less known that interfering beams that have very large optical path differences that yield unobservable intensity modulation, give a recombined beam possessing a spectral distribution having periodic minima. Spectral modulation occurs for optical path differences that far exceed the coherence length (sometimes referred to as photon length) of the interfering beams. This surprising statement is confirmed by experiments. Alford and Gold (1958) used a visible light source and found spectral modulation for an optical path difference of 64.2 meters, far exceeding the coherence length of their white-light unfiltered source. Cielo, Brochu & Delisle (1975) confirm spectral modulations in the spectrum of a visible light source for path differences as large as 300 meters. Mandel (1962) gives a full theoretical justification, while Givens (1961) gives a less rigorous but easier to follow physical explanation. While the original Alford and Gold experiment used pulsed sources, Givens (1961) predicted that continuous sources should also be spectrally modulated. This has been experimentally confirmed by Basano & Ottonello (2000).

The basic theory of the spectral modulation can be understood from the shift theorem. Consider an interferometer giving an optical path difference $c\tau$ into which one sends a pulse $V(t)$ having frequency spectrum $G(\omega)$ given by the Fourier transform of $V(t)$. The pulse could also be the wavepacket representing a single photon. The shift theorem gives the frequency spectrum of a pulse shifted in time by $\tau$ as

$$G'(\tau, \omega) = e^{-i\omega\tau} G(\omega). \tag{1}$$

The superposition of two interfering pulses shifted by $\pm \tau/2$ yields the frequency spectrum using the Fourier transform

$$H(\omega) = 1/(2\pi)^{1/2} \int_{-\infty}^{+\infty} [V(t+\tau/2) + V(t-\tau/2)] e^{-i\omega t} dt \ . \tag{2}$$

Applying the shift theorem (Eq. 1) to Eq. 2 we then obtain the well-known converse modulation theorem.

$$H(\omega) = (e^{i\omega\tau/2} + e^{-i\omega\tau/2}) G(\omega) = 2\cos(\omega\tau/2) G(\omega) \ , \tag{3}$$

Equation 3 shows that, after going through the interferometer, the spectrum of the source $G(\omega)$ is now modulated by a $cos(\omega\tau/2)$ term . These textbook results (Bracewell 1986) give the bases of the method by which we can determine the time delay $\tau$ between the two beams of an interferometer.

While the experimental set up used by Alford & Gold (1957) is essentially a Young slit interferometer that uses a white light pulsed source, it has the peculiar feature that it does not measure directly the spectral modulation predicted by equation 3 in the optical spectrum $H(\omega)$. Consider that Equation 3 predicts that the minima are spaced by

$$\Delta\omega = 2\pi\Delta n/\tau, \qquad (4)$$

where $n$ is an integer number. The optical path difference in the AG experiment (64.2 meters) thus gives a tiny frequency spacing $\Delta\omega$ impossible to measure with a spectrograph. Alford & Gold (1957) instead measure the minima in the beat spectrum detected in the current fluctuations of a quadratic detector (a photomultiplier in their experimental setup). Likewise, the proposed technique to observe gravitational lenses would measure the current $I(t)$ and detect the spectral minima of the fluctuations of $I(t)$ that appear at much lower frequencies than those of $H(\omega)$.

Givens (1961) shows that the frequency spectrum $I(\omega')$ of the output current $I(t)$ measured by the photomultiplier is given by

$$I(\omega') = (2\pi)^{1/2} \cos(\omega'\tau/2) \int a(\phi+\omega')a^*(\phi)G(\phi+\omega')G^*(\phi)d\phi \qquad (5)$$

where $a(\omega)$ is the bandpass of the detector, $\omega'$ is the beat frequency and the asterisk indicates complex conjugation. If we approximate $a(\omega)G(\omega)$ by a Gaussian having dispersion $\sigma$, Eq. 5 gives

$$I(\omega') = K\cos(\omega'\tau/2)\exp[-\omega'^2/(2\sigma^2)]. \qquad (6)$$

All the constants occurring after integration are grouped in K. The discussion leading to Eq. 6 is borrowed from Alford & Gold (1957) and Givens (1961) and summarized here for convenience. Consider that $\omega'$ is a beat frequency $\omega' = \delta\omega$ given by the difference $\delta\omega$ between any two frequencies $\omega$ and $\omega + \delta\omega$ in the primary spectrum (Givens 1961) and that $\omega'$ is obtained by tunable equipment placed after the detector: One can vary $\omega'$ while the frequency of observation $\omega$ remains constant. Equation 6 therefore predicts that $I(\omega')$ is spectrally modulated with periods $= 2\tau/(2n+1)$ and can be measured at a convenient low beat frequency $\omega'$ of our choice. Spectral modulation, which depends on the time delay $\tau$, therefore occurs at arbitrarily low frequencies $\omega'$ well outside the spectral bandpass of the light seen by the detector. As discussed at length by Givens (1961) and Mandel (1962), the AG effect allows one to measure arbitrarily long values of $\tau$, far larger than the coherence time given by the bandpass $\Delta\omega$ of observation.

To intuitively understand Equation 6, let us follow the discussion in Givens (1961). Let us first consider that the instantaneous current fluctuations $I(t)$ are due to wave beats among all the frequencies in the primary light beam, much as wave beats modulate the carrier frequency in a radio detector (Givens 1961). In the AG effect, the spectral modulation of the beat spectrum is then simply a consequence of the spectral modulation of the primary spectrum given by Equation 3. This can be understood by considering that if some high frequency waves of the primary spectrum are missing, there can be no beating with these missing waves, resulting in minima, and therefore spectral modulation, of the beat spectrum.

Consider next that the minima in the primary spectrum are a function of the optical path difference (Eq. 3 and 4). If the range of optical path differences in an extended source $A$ is sufficiently large, the minima in the primary spectra from all the small sections $dA$ will be spread over a wide range of frequencies. Consequently, if we add the spectrally modulated contributions from the different sections of an extended source, the sum will give negligible spectral modulation in the integrated primary spectrum. Prima facie, one may expect that there would be no modulation in the beat spectrum of the source. *However, this is only the case if the waves originating from independent microscopic light sources (e.g. atoms) beat among each other, since the small section dA are obviously independent.* We shall show below that this is not the case under the conditions of the experiment and that, in practice, a microscopic source only beats with itself. Consequently we must add beat spectra and not primary spectra. This article is based on this fact.

### 3. FREQUENCY DEPENDENCE OF THE SPATIAL CORRELATION COEFFICIENT

We shall show that the spatial coherence function that applies to the proposed interferometer depends on the beat frequency $\omega'$. To do this we shall first prove the fundamental fact that, in practice, a microscopic source only beats with itself. We will then proceed to write the spatial coherence function $\gamma(0)$ as a function of the beat frequency $\omega'$. The technique is based on this dependence on $\omega'$.

The mathematical treatment in the section 2 above implicitly assumes a perfectly spatially coherent source. The more rigorous treatment of Mandel (1962) uses formal statistical optics methods and includes spatial coherence effects. After a lengthy mathematical treatment he then shows that the contrast in the spectral modulation of the beat spectrum depends on $\gamma(0)$. Mandel (1962) assumes that the coherence function, which measures the visibility of fringes in classical interferometry as well as the visibility of the spectral modulation in spectral interferometry, can be written as the product $\gamma(0)\gamma(\tau)$ of the two terms commonly called spatial and temporal coherence functions. This commonly made assumption, while practical for computational purposes in many instances, is artificial since physically there only is a coherence function given by the cross-correlation between the advanced and retarded electromagnetic fields. In practice $\gamma(0)$ must be evaluated at a specific frequency. For quasimonochromatic light, typical of astronomical observations, $\gamma(0)$ is commonly evaluated at the central frequency $\omega_0$ of the bandpass of observation ( Klein & Furtak 1986). However, in the AG effect the detector observes at a frequency $\omega$ (e.g. a few $10^{15}$ Hz) but the measurements are made at the much lower beat frequency $\omega'$ (e.g. a few *MHz*) of the current fluctuations of a quadratic detector. The question that now arises is which frequency applies in our case. It cannot be answered with the formalism used by Mandel (1962) because he starts with the assumption of cross-spectral purity (Mandel 1961) that allows him to write the correlation coefficient as the product $\gamma(0)\,\gamma(\tau)$, carrying $\gamma(0)$ untouched down to the final equations. We shall instead reexamine the problem from first principles, starting from the superposition principle of electromagnetic waves. Classical electromagnetic theory is valid for our purpose provided the source is sufficiently intense. Mandel (1962) discusses the effect of photon shot noise on the AG effect (see also section 3.3 below).

### 3.1    Proof that, in practice, a microscopic source only beats with itself.

This section proves the fundamental assumption on which the technique is based. Consider a macroscopic region of the source considerably larger than the coherence area. It emits electromagnetic radiation given by the sum of the contributions of a very large number *n* of independent microscopic sources that have random phases $\phi_j(t)$, where the time dependence indicates that the phases can change with time. For example, in an atomic source, the time variation is caused by collisions with other atoms that randomize the phase. For simplicity, let us only consider beats between two frequencies $\omega_l$ and $\omega_k$ in the primary spectrum. The mathematics can readily be extended to beats among an infinite number of frequencies by summing over all the indices *k* and *l* in Equation 7. We can now follow an analysis analogous to the one carried out for classical interferometry (e.g. chapter 3 in Loudon, 1983) to show that, under the conditions of the experiment, a microscopic source in practice only interfere with itself. Using the superposition principle we have that the total electromagnetic field is given by

$$E(t) = E_0 e^{-i\omega_k t}(\sum_{j}^{n} e^{-i\phi(t)_k^j}) + E_0 e^{-i\omega_l t}(\sum_{j}^{n} e^{-i\phi(t)_l^j}), \qquad (7)$$

where the sum is extended over a very large number *n* of microscopic sources. The intensity $I(t)$ measured by the detector is then given by the time average of the product of Equation 7 with its complex conjugate

$$\begin{aligned}
I(t) = & E_0^2 <(\sum_{j}^{n} e^{-i\phi(t)_k^j})(\sum_{j}^{n} e^{i\phi(t)_k^j})> + E_0^2 <(\sum_{j}^{n} e^{-i\phi(t)_l^j})(\sum_{j}^{n} e^{i\phi(t)_l^j})> \\
& + E_0^2 < e^{-i(\omega_k-\omega_l)t}(\sum_{j}^{n} e^{-i\phi(t)_k^j})(\sum_{j}^{n} e^{i\phi(t)_l^j})> \qquad (8) \\
& + E_0^2 < e^{-i(\omega_l-\omega_k)t}(\sum_{j}^{n} e^{-i\phi(t)_l^j})(\sum_{j}^{n} e^{i\phi(t)_k^j})>
\end{aligned}$$

where the brackets $<\ >$ signify a time average over the integration time of the equipment that measures the current and where trivial multiplicative constants (e.g. quantum efficiency of the detector) have been neglected. This time is large with respect to the period of variation of the interfering waves but short compared to the period of the beat waves. Because the statistical properties of typical astronomical sources are stationary, we can invoke the ergodic theorem, as used in chapter 3 of Loudon (1983), to replace the time average by the statistical average. Following the reasoning of chapter 3 of Loudon (1983), we see that after multiplying out the brackets, because the phase angles of the wave trains from different sources have different random values, the cross-terms give a zero average contribution. The products of the remaining terms give

$$I(t) = nE_0^2 + nE_0^2 + nE_0^2(e^{-i(\omega_k - \omega_l)t} + e^{-i(\omega_l - \omega_k)t}) = 2nE_0^2\{1 + \cos[(\omega_k - \omega_l)t]\} = 2nE_0^2\{1 + \cos(\omega't)\}$$
, (9)

where $\omega_k - \omega_l = \omega'$ is the beat frequency, $n$ is the total number of microscopic sources and the term it multiplies is the contribution from a single microscopic source. Because the $n$ factor in Equation 9 multiplies the contribution from an individual microscopic source, Equation 9 shows that, in practice, a microscopic source only beats with itself. Consequently, the integrated beat spectrum is given by the sum over the beat spectra of individual microscopic sources. *This conclusion is fundamental for it will allow us to add the contributions of individual beat spectra in section 3.2.*

Under particular conditions, different microscopic sources can actually beat among themselves (Forrester 1961, Forrester, Gudmunden & Johnson 1955). However, this apparent contradiction with our demonstration in this section is analogous to the apparent contradiction that occurs in the demonstration that atoms only interfere with themselves in classical interferometry (Klein & Furtak 1986), while it is known that independent sources can interfere among themselves under particular coherence conditions (Magyar & Mandel 1963). The beats observed by Forrester, Gudmunden & Johnson (1955) only occur within small coherence regions (i.e. the subregions where Eq. 10 applies). Forrester (1961) shows that adding the contributions from these regions destroys the beats, in agreement with our discussion in this section. While the definition of a coherence region made by Forrester (1961) uses small areas on the detector and is somewhat different from the more familiar definition in terms of areas on the sources, the two are equivalent (Forrester 1956).

3.2 **The Spatial Coherence Function**

Consider an extended source observed through an interferometer that has a large optical path difference such that the spectrum of the output beam has the spectral modulation predicted by Eq. 3. The minima in this primary spectrum are obliterated by adding contributions from independent microscopic sources having different optical path differences. If independent microscopic sources could beat among themselves, the minima in the beat spectrum would then obviously be obliterated as well. However, we have just shown that, in practice, the microscopic sources do not beat among themselves; consequently only beat waves from independent microscopic sources do add up to form the beat spectrum. Because electromagnetic waves from independent sub-regions on the surface of a macroscopic source obviously come from independent sources, it is then obvious that waves from these subregions will not beat among themselves.

Consequently, what degrades the spectral modulation of the beat spectrum from an extended macroscopic source is the superposition of the beat spectra coming from its subregions. We could therefore have an extended macroscopic source with a primary spectrum that has unobservable modulation but yet has a measurable modulation of its beat spectrum because waves from different subregions, while obliterating the modulation in the primary spectrum, do not beat among themselves.

We can now proceed and quantify the effect in terms of familiar parameters. We shall follow a discussion similar to the discussion of spatial coherence in Klein & Furtak (1986), who use a Young interferometer model having slits separated by the distance $a$, and therefore shall use similar notation. The discussion in Klein & Furtak (1986) follows the standard treatment of spatial coherence that can be found in many optics textbooks. The spectral contribution of the spectrally modulated beat spectrum from a region of the source small enough that spatial coherence effects are negligible can be written as

$$D(\omega') = S(\theta)\Psi(\omega')[1+\cos(\omega'\tau)], \qquad (10)$$

where $\psi(\omega')$ is the spectral density of the unmodulated beat spectrum and $S(\theta)$ the intensity at angle $\theta$. Equation 10 approximates well the shape of the beat spectrum that is actually observed (Alford & Gold 1958). For simplicity, like in Klein & Furtak (1986), we only assume a radial dependence of the surface intensity on $\theta$. Considering that Eq. 10 is similar to the equations used in classical coherence theory (e.g. chapter 8 in Klein & Furtak 1986) to express intensity contrast for temporal and spatial coherence in classical interferometry, we can follow a similar mathematical treatment to derive the equations that express spectral contrast for an extended source by only changing symbols. Therefore, in the section 8.3 of Klein & Furtak (1986) that considers spatial coherence we simply replace $2\pi\nu_0$ by $\omega'$. By doing this, we follow the identical procedure adopted in Klein & Furtak (1986), where spatial coherence theory is obtained by simply changing symbols in the mathematical analysis of temporal coherence theory.

Because equation 9 shows that electromagnetic fields from different microscopic sources, and therefore different regions of the source, do not interfere, we can obtain the spectral density $D(\omega')$ for an extended source having a surface intensity distribution $S(\theta) = S_0 i(\theta)$ by integrating Eq. 10 over the surface of the source, writing (see Klein & Furtak 1986 for intermediate steps leading to Eq. 11)

$$D(\omega') = S_0\Psi(\omega')\int_{-\infty}^{\infty}\{i(\theta)+i(\theta)\cos[\omega'(\tau-a\theta/c)]\}d\theta, \qquad (11)$$

Where $i(\theta)$ is the normalized angular distribution function, $\tau$ is the mean optical path difference introduced by the interferometer which depends on $a$ and therefore the mass of the lensing object, and $S_0$ the total flux density. For simplicity we have assumed that $\psi(\omega')$ is constant over the entire surface. In our Young interferometer model of gravitational lensing, the impact parameter $a$ quantifies the time delays among the two images of a lensed source and has therefore the same effect as the separation $a$ between the slits of the Young interferometer. The impact parameter *a* therefore depends on the mass of the lensing object. The second term in the integral can readily be identified as the normalized correlation function. Following Klein & Furtak (1986), after switching to the complex notation in the integral, we can then write this second term within the integral in Eq. 11 as

$$\int_{-\infty}^{\infty} i(\theta)\cos[\omega'(\tau - a\theta/c)]d\theta = \gamma(\tau, a, \omega') = \gamma(a, \omega')\cos(\omega'\tau), \qquad (12)$$

with

$$\gamma(a, \omega') = \text{Re}[\int_{-\infty}^{\infty} i(\theta)e^{-i\omega'a\theta/c}d\theta], \qquad (13)$$

where Re[] signifies that we take the real part of the complex integral and $\gamma(a,\omega')$ is therefore the Fourier transform of the normalized intensity distribution $i(\theta)$ and can readily be identified as the spatial correlation function $\gamma(0)$ evaluated at frequency $\omega'$. Equation 11 can then be written as:

$$D(\omega') = S_0\psi(\omega')[1 + \gamma(a, \omega')\cos(\omega'\tau)]. \qquad (14)$$

Equations 13 and 14 resemble the familiar equations of classical interferometry with the outstanding differences that they express a spectral modulation, rather than intensity modulation and depend on a beat frequency ω'. We can see, from Eq. 14, that $\gamma(a,\omega')$ quantifies the visibility of the spectral modulation with the visibility given by

$$V(a, \omega') = (D\max - D\min)/(D\max + D\min) = \gamma(a, \omega'). \qquad (15)$$

## 4. DISCUSSION

Equations 14 and 15 quantify the loss of contrast in the spectral modulation of the beat spectrum of a gravitational lens. The visibility of the spectral modulation depends on the Fourier transform of the intensity distribution of the source evaluated at the beat frequency $\omega'$ (Eq. 13). To gain an insight into the mathematics and physics of the situation, let us approximate the lensed object with a simple model consisting of a disk having uniform intensity ( *i(θ) = 1.0* ) and angular diameter *Δθ*. The mathematics of the problem is then the same as the well-known mathematics of the Fourier transform of the pupil of a circular aperture and has the same solution. Eq. 13 therefore gives, as its Fourier transform, a solution *V(a,ω')* that is the absolute value of a normalized first-order Bessel function, having a maximum = 1.0 at  *ω' = 0.0* and a first zero at

$$\omega' a\Delta\theta/(2\pi c) = 1.22. \qquad (16)$$

By observing the profile of *V(a,ω')* (Figure 2), we can gain an insight on how the observations of a gravitational lens would be carried out. Because of the symmetry between *ω'* and *a* in Eq. 13 and Eq. 16, we see that changing frequency *ω'* is equivalent to changing the spacing *a* in a Young interferometer. In the gravitational lens the impact parameter *a*, analogous to the spacing between the slits of the Young interferometer, depends on the mass of the lensing object and is thus constant but we are free to increase $\omega'$. To find a lensed object one would vary $\omega'$ and therefore $\gamma(a,\omega')$ and $V(a,\omega')$ (Equations. 13, 15 and Fig. 2) until the spectral modulation given by Eq. 14 becomes apparent. The task would obviously be facilitated if one had a mass range target for the lensing object since the mass of the lensing object is a key parameter for the impact *a*. For a given mass there is a range of values of *a* that depend on the geometry of the lensing event (Refsdal 1964, Press & Gunn 1973).

At this point, let us emphasize that the frequency that changes is NOT $\omega$, the frequency of observation, but the beat frequency $\omega'$. This distinction is fundamental. One could, for example, observe at a frequency of a few GHz, obtain the beat spectrum $D(\omega')$ by measuring the spectral distribution of the fluctuations of the current of the detector at frequencies $\omega'$ of the order of a few MHz and simply determine how the visibility of the spectral minima of $D(\omega')$ varies with $\omega'$. In the AG experiment, the beat frequency is simply varied by changing the frequency of a short-wave receiver.

Our discussion has so far neglected the effect of noise. Since a detailed discussion of the signal to noise ratio depends on the instruments and techniques used, which themselves depend on the region of the electromagnetic spectrum, it is beyond the scope of this paper. We can however make estimates of the signal to noise ratio based on photon shot noise. Using the work of Mandel (1962) and Purcell (1956), Borra (1997) gives the ratio *R* between the wave-interaction and shot noise terms of the spectral densities

$$R \leq \delta = \alpha \bar{I} 2\pi / \Delta\omega \quad , \tag{17}$$

where $\delta$ is the degeneracy parameter (Mandel 1962), $\alpha \bar{I}$ is the average count rate, $\Delta\omega$ the bandpass of observation and $2\pi/\Delta\omega$ gives the usual estimate of the coherence time approximated by the inverse of the width of the spectral bandpass. The parameter $\delta$ therefore gives the average count rates inside the bandpass. Equation 17 shows that shot-noise dominates whenever $\delta<<1$, which is when the counting rate is substantially below one count per coherence time interval $2\pi/\Delta\omega$, while the spectral modulation becomes easier to detect if $\delta>1$

Equation 17 predicts that, for a given flux, the degeneracy parameter $\delta$ is considerably larger in the radio region than in the optical since the energy of a photon, $h\omega/2\pi$, is smaller and the coherence times $2\pi/\Delta\omega$ larger at lower frequencies. Since the degeneracy parameter $\delta$ increases with decreasing frequency, the effect will be easier to detect if one observes at low frequencies. Assuming a 100-m diameter radio-telescope observing at *1 GHz* and $\alpha=0.5$, Eq. 15 gives $\delta=1$ for a *4 milliJansky* source (Borra 1997). Note that $\delta$ does not depend on the width of the bandpass $\Delta\omega$ since the increase in counts is proportional to $\Delta\omega$, and is therefore compensated by the $1/\Delta\omega$ factor in Eq. 17.

The degeneracy parameter $\delta$, which quantifies the importance of photon shot noise in wave interaction physics, is unfamiliar to most Astronomers and it is therefore worthwhile to carry out a brief discussion on the limiting fluxes using astronomical units. Let us first note that the issue of photon noise on wave interaction phenomena is not a simple one, as can be appreciated by the controversies that surrounded the introduction of Intensity interferometry in the 1950s (Brannen & Ferguson 1956, Hanbury Brown Twiss 1956, Purcell 1956). Purcell (1956) recommends caution in using wave packets to represent photons. Chapter 9 in Goodman (1985) examines at length the fundamental limits in photoelectric detection of light and gives relations that can be used to obtain estimates of limiting fluxes. Keeping in mind that the exact relations will depend on the particular frequency and techniques used, we can start from the *4 milliJansky* flux, given above, that yields $\delta = 1$ for a source observed with a 100-m radiotelescope observing at 1 GHz. The relation between $\delta$ and limiting flux depends on whether the technique uses first-order or second order correlations. A first order correlation applies, for example, if one simply uses a mathematical Fourier analysis of the output current *I(t)*; while a second order correlation would apply if one uses a technique similar to the one used in the original AG experiment (Borra1997). Assuming the worse-case scenario of second-order correlations we can use equation 9-5-20 in chapter 9 of Goodman (1985) that gives the integration time *t* for second order correlations as function of the signal to noise ratio *S/N* averaged over many counting intervals $t_0$ as

$$t/t_0 = (S/N)^2 /(\delta^2 V^4), \qquad (18)$$

where $\delta$ is the degeneracy parameter and *V* the visibility parameter (Eq. 15). Assuming the values also assumed by Goodman (1985) for the Hanbury-Brown-Twiss experiment ($\delta = 10^{-3}$, $V = 0.1$, $t_0 = 10^{-7}$ second and $S/N = 10$) we then obtain a measurement time of $10^5$ seconds (28 hours) to reach *4 microJanskys* with a 100-m radiotelescope observing at *1 GHz*. However, this flux limit is almost certainly optimistic since it would be even lower than the flux limit obtained for ordinary flux measurements. This is because Eq. 18 only takes into account photon shot noise and neglects other sources of noise like system noise and confusion noise that would be present in actual astronomical observations. It therefore only signifies that wave-interaction effects dominate and that photon noise is not the limiting factor. This suggests that, at 1 GHz, one should probably be able to observe to limiting fluxes similar to those obtained with ordinary flux measurements. This statement is not surprising since the intensity interferometer can observe to flux limits comparable to those of the Michelson interferometer in the radio region (Twiss 1969). Twiss (1969) discusses criteria as to when an intensity interferometer is competitive with a Michelson interferometer. Limiting fluxes with a telescope having a different diameter can be obtained using the fact that the degeneracy parameter $\delta$ can be interpreted as the average number of photon counts that occur in a single coherence time: Consequently $\delta$ is proportional to the collecting area of the telescope.

To obtain exact signal to noise versus flux relations one needs to know details about the instrumental systems (e.g. correlation order, frequency of observation and detector) that will actually be used. The signal to noise ratio could be increased with more sophisticated data analysis techniques. For example, one could use the fact that Eq. 14 predicts multiple minima separated by $\delta\omega' = 2\pi/\tau$. One could use fitting techniques to extract the signal from a noisy continuum.

## 5. CONCLUSION

The fact that the visibility of the spectral modulation of the current fluctuations depends on a spatial correlation function $\gamma(0)$ which is evaluated at beat frequencies $\omega'$ opens up applications to gravitational lensing. Because we are free to choose $\omega'$, there is the important consequence that we are no longer limited by the angular size of the lensed object nor the mass of the lensing object. This can be understood by looking at Figure 2, which shows the contrast function that quantifies the visibility of the spectral modulation. Because the independent variable in that figure is $\omega'a\Delta\theta/(2\pi c)$, we see that by choosing $\omega'$ we can position ourselves in the location in the figure that best suits our purposes, irrespective of the values of the impact parameter $a$, which depends on the mass of the lensing object and $\Delta\theta$, the angular size of the source. This signifies that the technique does not suffer from the strong limitations that it would suffer if $\gamma(0)$ depended on the frequency of observation (Borra 1997).

Numerous articles discuss applications of gravitational lensing. A detailed discussion of the application of the technique to gravitational lensing is beyond the scope of this article. We shall instead consider simplified equations to give order of magnitude estimates. The equation that gives the time delays (Refsdal 1964b, Press & Gunn 1973, Spillar 1993) depends not only on the mass and mass distribution of the lens but also on the geometry of the event (e.g. distances between source, lensing object and observer) and cosmological parameters. For discussion purposes we shall use the simple relations in Press & Gunn (1973) who give an approximate formula for the separation $\theta$ for images of comparable brightness of a source at $z_{source} = 2$ for $H0 = 60\ kms^{-1}Mpc^{-1}$ lensed by a point-like mass, as well as, assuming a simple geometry where the lensing object is located at 1/3 of the distance between the source and the observer, another approximate formula for the time delay as a function of $\theta$. We thus obtain a relation, approximately valid for the above mentioned parameters,

$$\tau = 3.4 \times 10^{-5} M\ \text{sec}, \qquad (19)$$

where the time delay $\tau$ is in seconds and the mass $M$ is in solar units. Spillar (1993) estimates time delays valid for a Euclidean universe and a geometry where the distance between the lensing object and the observer is considerably smaller than the distance between the source and the observer. He obtains time delays comparable to those obtained with Eq. 19. For solar mass objects Eq. 19 yield $\tau = 3.4\ 10^{-5}$ seconds and minima in $D(\omega')$ separated by $\delta\omega' = 2\pi/\tau = 1.8\ 10^5\ Hz$ (Eq. 14). Lenses having masses of the order of a solar mass or a small fraction of a solar mass give thus time delays that are within orders of magnitude of the time delays measured in the laboratory and could therefore be readily measured. We could observe them at beat frequencies of the order of several MHz. We can see that the technique could readily identify and measure lensing events that are unobservable with current techniques.

According to the cold dark matter models, a large number of dark subhalos should be located within the halo of each galaxy. Millilensing from these halos is another example of a situation where the separation among the components is too small to be detected by conventional techniques. For masses between $10^5$ and $10^{11}$ solar masses equation 19 predicts values of $\tau$ between 3.4 and $3.4\ 10^6$ seconds (39 days). These are substantially larger than the values of $\tau$ that have been measured in the laboratory with the AG effect but there is no reason why they should not be measurable.

We can also use a simple estimate of the impact parameter $a$ from Press & Gunn 1973).

$$a = 10^{17} M^{1/2}\ \text{cm}, \qquad (20)$$

where $a$ is in cm and the mass $M$ is in solar units. Using Eq. 20, we can approximate Eq. 16 by

$$5 \times 10^5 M^{1/2} \omega' \Delta\theta = 1.22. \qquad (21)$$

Equation 21 gives useful order of magnitude estimates, within the limitations of the model (e.g. a uniform disk of angular diameter $\Delta\theta$).

If we assume for estimate purposes a simple Euclidean geometry and a source having a diameter of 1 astronomical unit at a distance of 1 Gpc, we obtain $\Delta\theta = 5 \times 10^{-15}$. Eq. 21 therefore predicts, for a solar mass lens, that the first minimum in Fig. 2 would occur at a beat frequency of 80 MHz. The spectral modulation would be observable at the beat frequencies of a few MHz predicted by Eq. 19. Since Eq. 21 predicts that $\omega'\Delta\theta$ scales as $M^{-1/2}$, we see that a $10^6$ solar mass object would have a first zero at 80 KHz and could therefore be observable at frequencies of several KHz. All of these estimates are only approximations but give order of magnitude estimates that show that the effect should be observable.

Besides finding unresolved lenses, the technique could also be used to obtain information on the lensing event and the source itself. Consider that $\gamma(a,\omega')$ given in Eq. 13 is the Fourier transform of the intensity distribution $i(\theta)$ of the source. The structure of the source $i(\theta)$ could then be obtained by the inverse Fourier transform of $\gamma(a,\omega')$ where the variable is the beat frequency $\omega'$. As a simple illustration of this application, let us model the source by a disk of uniform intensity so that its Fourier transform is the absolute value of a first order Bessell function. The contrast function is then given by Figure 2. To determine the diameter of the lensed object one would first need the impact parameter *a* from the time delay $\tau$ obtained from the minima in the spectral modulation (Eq. 14). The impact parameter could either be obtained from detailed modeling or approximate relations (e.g. Eq. 19 and Eq. 20). Fitting the observed contrast function $V(\omega',a)$ to the theoretical function shown in Fig. 2 would determine $\gamma(\omega',a) = 0.0$ (making sure it is the first zero) thus obtaining the diameter $\Delta\theta$ of the disk from Eq. 16 .

The determination of cosmological time delays is perhaps the most interesting application of the technique. It has been proposed that time delays can be used to determine cosmological parameters (Refsdal 1964b, Blandford & Naryan 1992). Time delays using time variations of luminosity in the source are difficult to measure but would be far easier to measure accurately with the technique.

The major inconvenience of the technique is that it needs strong sources, as can be understood from Mandel's (1962) discussion of the effect of shot noise (see also section 4). This inconvenience is however mitigated by the fact that, like in the intensity interferometer experiment (Hanbury Brown 1968), the technique can work with inexpensive primary mirrors, having surface qualities much lower than those required for conventional telescopes (Borra 1997). This is perhaps the most intriguing feature of the technique since one could build specialized inexpensive telescopes using very large primary mirrors having optical quality vastly more relaxed than those for conventional telescopes to gather the large fluxes that would allow us to minimize the degeneracy handicap (Eq. 15).

Borra (1997) briefly discusses the effects of interstellar scintillation and gravitational wave background.

Our analysis used the very simple model commonly used in gravitational lensing that only approximates the actual physics. For example, we have assumed a simple two-beam geometry, while gravitational lenses can have multiple beams of unequal intensities. The uniform disk model is another obvious simplification. These can all be modeled and the modeling can be used to derive useful parameters (e.g. surface structure). A discussion of this is beyond the scope of the present article (see Borra, 1997 for further discussion).

The predictions made in this work remain to be experimentally confirmed. However, consider that we follow a mathematical analysis that starts from the same first principles and then follows a mathematical treatment identical to the analyses used to describe temporal and spatial coherence in classical interferometry. Since both are fully confirmed by experiments, we should expect that the present analysis is valid as well.

Finally, note that our discussion does not apply to the interferometric measurements of time delays also discussed by Borra (1997). In that case, the correlation function must be evaluated at the frequency of observation $\omega$.

ACKNOWLEDGEMENTS

This research has been supported by the Natural Sciences and Engineering Research Council of Canada.## REFERENCES

Alford, W. P., Gold, A.,1958, Am.J.Phys. , 26, 481
Basano, L., Ottonello, P., 2000, Am.J.Phys., 68, 325
Blandford, R.D., Naryan, R., 1992, ARA&A, 20, 311
Brannen, E., Ferguson, H.I.S , Nature, 178, 481
Borra, E.F., 1997, MNRAS, 289, 660
Bracewell, R.N., 1986, The Fourier Transform and its Applications, McGraw-Hill, New York.
Cielo, P.,  Brochu ,M., Delisle, C., 1975, Can.J.Phys., 53, 1047
Givens., M. P.,1961, J.Opt.Soc.Am., 51, 1030
Goodman, J.W., 1985, Statistical Optics. John Wiley & Sons. New York.
Forrester, A.T., 1961, J.Opt.Soc.Am., 51, 253
Forrester, A.T., 1956 Am.J.Phys., 24,1,92
Forrester, A.T., Gudmunden, R.A., Johnson, P.O., 1955, Phys. Rev. 99, 1691
Hanbury Brown, R., Twiss, R.Q., 1956, Nature, 178, 1447
Hanbury Brown, R., 1968, ARA&A, 6, 13
Klein, M. V., Furtak, T. E., 1986, Optics, John Wiley & Sons, New York
Loudon, R., 1983, The Quantum Theory of Light, Oxford Univ. Press, Oxford
Magyar, G., Mandel, L., 1963, Nature, 198, 255
Mandel L., 1961, J.Opt.Soc.Am., 51, 1342
Mandel L., 1962,  J.Opt.Soc.Am., 52, 1335
Peterson J.B., Falk, T., 1991, ApJ, 374, L5
Purcell, E.M., 1956, Nature, 178, 1449
Press, H.W., Gunn, J.E., 1973, ApJ, 185, 397
Refsdal, S., 1964, MNRAS, 128, 23
Refsdal, S., 1964b, MNRAS, 128, 307
Spillar, E. J., 1993, ApJ,  403, 20
Twiss, R.Q. Optica Acta, 1969, 16, 423
16

FIGURE CAPTIONS.

Figure 1: It shows a layout of the gravitational lens interferometer. The gravitational lens is modeled by a Young interferometer having equal strength beams.

Figure 2: It shows the contrast function $V(a,\omega')$ that quantifies the visibility of the spectral modulation.

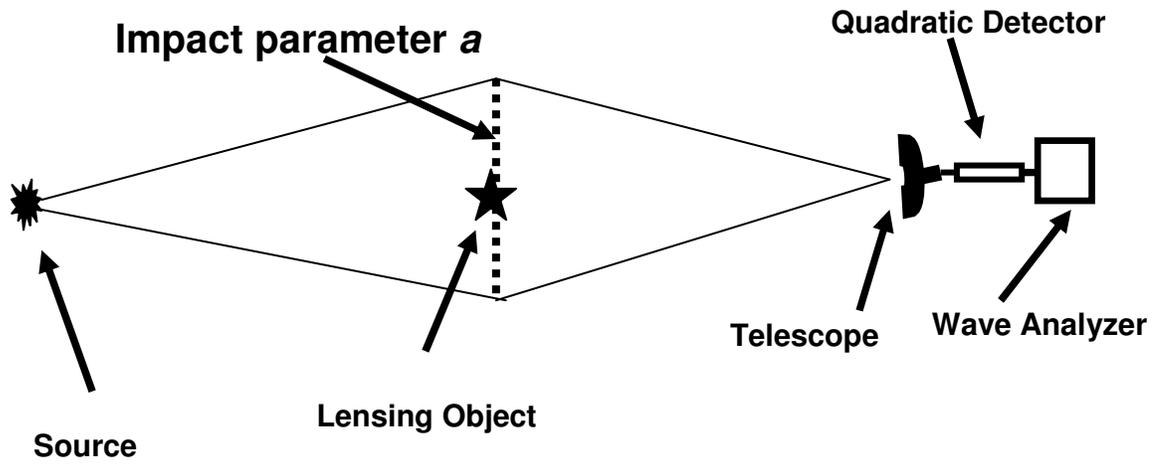

FIGURE 1

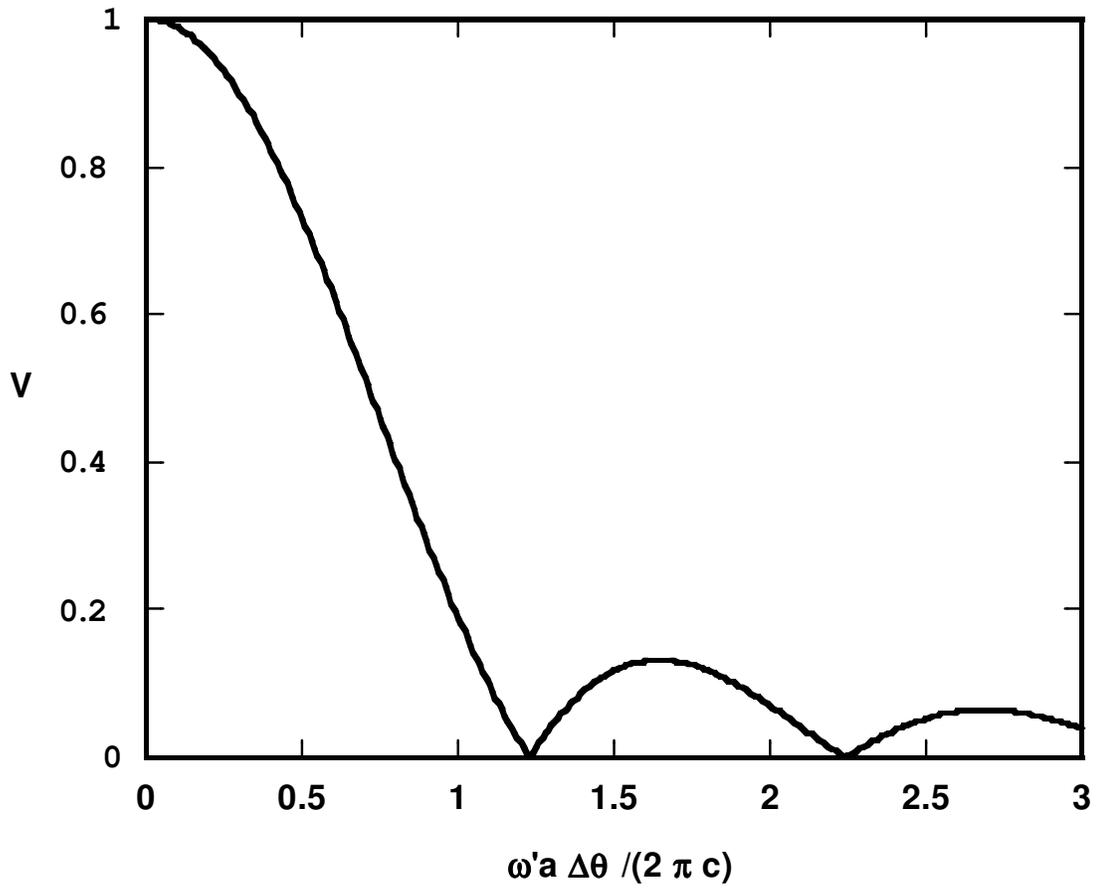

FIGURE 2